# Security - a perpetual war: lessons from nature

Wojciech Mazurczyk and Elżbieta Rzeszutko

Warsaw University of Technology, Institute of Telecommunications

Warsaw, Poland, 00-665, Nowowiejska 15/19

Email: wmazurczyk@tele.pw.edu.pl, E.Rzeszutko@tele.pw.edu.pl

**Abstract.** For ages people have sought inspiration in nature. Biomimicry has been the propelling power of such inventions, like Velcro tape or "cat's eyes" – retroreflective road marking. At the same time, scientists have been developing biologically inspired techniques: genetic algorithms, neural and sensor networks, etc. Although at a first glance there is no direct inspiration behind offensive and defensive techniques seen in the Internet and the patterns present in nature, closer inspection reveals many analogies between these two worlds. Botnets, DDoS (Distributed Denial of Service) attacks, IDS/IPSs (Intrusion Detection/Prevention Systems), and others, all employ strategies which very closely resemble actions undertaken by certain species of the kingdoms of living things. The main conclusion of the analysis is that security community should turn to nature in search of new offensive and defensive techniques for virtual world security.

## 1. Introduction

Nature is probably the most amazing and recognized invention machine on Earth with its about three billion years of experience in evolution by natural selection, genetic drift and mutations. Therefore it is not surprising that it has inspired inventors and researchers for ages. There are many examples that can be mentioned but some famous, modern cases include:

- **Velcro** (fabric hook and loop fastener) – created by close inspection of burdock burrs by the Swiss engineer George de Mestral after his hunting trips in Alps in 1941,
- **"Cat's eyes"** – retroreflective road marking invented by an English businessman – Percy Shaw – in 1933 and their inspiration was the shine reflecting from the eyes of a cat when it crossed the road while Shaw was driving at night.

- **Olympic swimmers' fast suit** fabric replicating shark's dermal denticles to reduce drag. The sharkskin swimsuits were first produced by Speedo company, and their use in the 2008 Summer Olympics helped breaking a few of the world records.

Nature's footprint is also present in the world of Information Technology, where there is an astounding number of computational bio-inspired techniques. Examples include genetic algorithms, neural networks, ant algorithms just to name a few. Networking technologies have also adopted some of nature's ways, which take form of swarm intelligence, artificial immune systems, sensor networks, etc. ([10], [11]).

Under close inspection a similar parallelism becomes visible in the field of digital security, but here the analogy is more a matter of coincidence, rather than the result of deliberate activity. Even though signature-based virus detection in antivirus software corresponds to immunization via vaccinations, the exact source of inspiration for this security tool remains ambiguous. We want to persuade the reader that both the defensive and offensive strategies happen to mimic the ongoing race of arms present on nature's battlefield of species. Many current network attacks, both complex and simple, that are known from the virtual world are in fact not novel at all. Worms, SPAM campaigns, (D)DoS ((Distributed) Denial of Service) attacks, etc., as well as defensive techniques like firewalls or ID/PSs (Intrusion Detection/Prevention Systems) all have their counterparts in the natural habitat. In the plant and animal kingdoms we can list many more examples where such techniques – both offensive and defensive – have been utilized for millions of years. Therefore even without direct inspiration from nature, the human kind is still unconsciously following its steps.

We discuss *selected* examples mainly from the plant and animal kingdoms matching the essence of network attacks and defences (Table I). However, it must be emphasized that for each of the illustrated techniques from the virtual world we can easily find a whole range of suitable examples from nature. It is worth noting that to the authors' best knowledge currently no such perspective that links virtual attacks and countermeasure techniques with the mechanisms for competition, dominance and defence observable in any of the Eukaryote domain kingdom's, has been presented in literature.

## 2. Offensive techniques – typical network attack scenario

To convince the reader about the analogies between IT security and the interaction scenarios among species within eukaryotes' kingdoms it will be most suitable to refer to the general yet simple network attack scenario.

**Table I** – Analogies between offensive/defensive techniques in the virtual world and the kingdoms of Eukaryote domain

| Essence/feature of offensive / defensive technique | IT Security example | Example from nature | Offensive (O) / Defensive (D) technique |
|---|---|---|---|
| Attracting the victim and fooling it to swallow the bait (fatal) | Phishing websites or emails | Anglerfish (*Lophius piscatorius*) | O |
| Disabling attack's countermeasures | Worms | Bolas Spiders (*Araneidae*) | O |
| Taking control over an entity to use it for own purpose | Botnet | *Ophiocordyceps unilateralis* (a pathogenic fungus) | O |
| Covert communication | Botnet | Philippine tarsier (*Tarsius syrichta*), Richardson's ground squirrel (*Urocitellus richardsonii*) | O |
| Generation of large volumes of unnecessary resources | SPAM | Small Balsam (*Impatiens parviflora*) | O |
| Legitimate entity prevented from using a resource | DDoS | Kudzu (*Pueraria montana*) | O |
| Attracting the victim to achieve a designated goal (not fatal) | Honeypot | Lady's-slipper orchid (*Cypripedium calceolus*) | D |
| Prevention of external threats | Firewalls | Hedgehog's spines, porcupine's quills, turtle's shells or acacia thorns | D |
| Ability to differentiate between what is welcome and what is not | Firewalls | Allelopathy phenomenon in a Mexican shrub (*Leucaena leucocephala*) | D |
| Intruder detection and attack prevention | IDS/IPS | Masked Birch Caterpillar – larvae of *Drepana arcuata* | D |

Typically, in many network attacks that happen every day in the Internet, we can distinguish the following phases:

**Phase 1:** Malware installation – users are fooled to run some malicious software (e.g. a virus or trojan) on their machines. This can be achieved by, e.g., means of phishing techniques, infected USB devices, etc.

**Phase 2:** Botnet creation – typically, the main aim of the malicious software is for the user's machine to become part of the botnet – a zombie machine.

**Phase 3:** Launching of an attack – botnets are mainly utilized to perform illegal activities, including conducting DDoS attacks, initiating SPAM campaigns, etc.

We refer to each of the network attack phases and for each of them we provide counterpart scenario taking place among the species of the kingdoms of Eukaryote domain. For the sake of simplicity we chose only one or two examples for each network offensive/defensive technique. However, it should be emphasised, that since species' interactions quite frequently are oriented toward competition or predation, there is a plethora of possible scenarios resembling the attack-defence events taking place in the world of digital security.

*2.1 Phase I – Malware installation*

In the Internet there is a whole variety of malicious software that can infect users' devices, among it: worms, viruses, spyware, trojans, etc. Moreover, there are many possible ways in which maleware can be installed on a victim's machine. The attack can be conducted, for example, by luring users with the aid of phising websites or emails, fake posts or other content on social networking websites. Other attack vectors include the use of infected removable devices or downloading and installing untrusted software from malicious websites.

In this subsection we show the analogies between worms, phising attacks and the corresponding events taking place in nature.

**Worms.** This is a type of self-propagating malware that can typically be characterized by two main features:

**(F1)** it utilizes exploits to install itself on a user's machine and

**(F2)** it is capable of disabling running security systems like firewalls or anti-virus software, etc.

The first characteristic is fairly commonly encountered in nature. Predators sometimes exploit the evolutionary perceptual bias of other organisms. If the prey is pre-evolved to respond to a sensory signal it becomes vulnerable to exploitation of its bias. A representative of the Arachnid class may serve as an example here. Bolas spiders produce signals that their prey has been pre-programmed to respond to. They create a viscous silk ball, which hangs on a thread – its effectiveness is increased, as the spider produces a chemical mirroring the female sex pheromone of its prey. Thus, by means of such aggressive mimicry, male moths are lured into the trap (F1).

Looking into the worm's other characteristic feature, i.e. the ability to disable defensive security systems, let us consider Arachnids, and the way in which spiders hunt their prey. Typically they utilize two attacking techniques to subdue their victims: the web they weave and the venom they inject through fangs. Both these techniques can be treated as means to disable prey's ability to defend itself or escape, as they lead to victim's paralysis or immobilisation (F2).

**Phishing.** Consider phishing websites. Their main goal is to masquerade a legitimate website and make the user give out his secret password, credit card number or similar. Thus, the essence of this attack technique is to attract the victim and fool him to swallow the bait.

Many predators among the species of animal and plant kingdoms have used this technique for ages. For example, let us consider Anglerfish (*Lophius Piscatorius*) which is sometimes referred to as the "sea-devil". The fish has eighty long filaments along the middle of its head, the most important being the longest one, terminating in a lappet, and is movable in every direction. The angler attracts other fish by means of its lure, to seize them with its enormous jaws as they approach.

*2.2 Phase II – Forming a botnet*

Once malware is planted on a computer, the first step towards establishing a botnet is completed. The user is unaware, but he loses the ability to control his machine – it becomes a "zombie". Many zombie computers form a botnet, controlled remotely by an attacker – the botherder – using a Command and Control (C&C) channel for issuing instructions regarding the actions to be undertaken by the zombie army. In this way the botherder can utilize the users' captured machines for his own purposes.

Moreover, recent trends revealed that, to provide more stealth, botnet C&C channels have started utilizing information hiding techniques called steganography. In September of 2011 a new worm called Duqu was found, whose general structural characteristics are similar to the infamous Stuxnet [1]. The most stunning intricacy in Duqu's functioning is its employment of special means for transferring of the obtained data through C&C channels to the malware's owners. The captured information is hidden in seemingly innocent pictures and traverses the global network as ordinary files, without raising any suspicion ([2], [3]). A similar functioning mechanism was found in a new variant of the Alureon malware [1], which was also discovered in the same period. Furthermore, in March 2014, ZBOT malware was identified to possess similar functionality, but here it was employed to hide a list of banks and financial institutions to which users' access will be monitored. Once the user visited any of the listed sites, the malware would try to steal his credentials [5].

In this subsection we show analogies for the following botnets' features:

**(F1)** Taking control over machines to utilize them for malicious purposes.

**(F2)** Providing covert communication in the Command and Control (C&C) channel to improve botnet's undetectability.

Despite the fact that gaining dominance over the behaviour of a living creature and steering its actions according to the dominator's will may sound somewhat fantastic, it does happen in the natural environment. *Ophiocordyceps unilateralis*, a species of fungi, is such an evolutionary marvel. This small fungus requires very specific conditions for development and reproduction. Its spores germinate only when attached to an ant's exoskeleton. The victim insect is slowly digested by enzymes secreted by the growing fungus, but before it is killed, its behaviour is altered by chemicals released by the *Ophiocordyceps* [8]. The zombie ants are, presumably by means of altering the functioning of their pheromone receptors, compelled to undertake a deadly journey. They climb plants, until they find a location characterised by an appropriate microclimate – such conditions happen to be found on the northern parts of the plant, where they are finally forced to dig their jaws into underside of the leaf's veins (to remain attached). What becomes ant's place of annihilation is the best location for the fungus to sprout, reproduce and release its spores, thus the cycle of life is completed. Looking back at botnets, we notice that here cycles are also present – a recruited zombie is quite frequently employed to play a role in other malicious activities, which can lead to recruitment of even more bots. Moreover it has been investigated that dead ophiocordyceps-infected ants often occur in close proximity to each other (even up to 26 ants/m$^2$) and form infected zones called ants' graveyards – these can be treated as special cases of zombie-ant networks. Any ant that will enter such a region will eventually become infected and will enlarge the number of recruited insects (F1).

The ability to communicate covertly is a precious evolutionary achievement, which can be used for both defensive and offensive purposes. The value of this skill has proven high among the *Philippine tarsiers* (*Tarsius syrichta*) which are small nocturnal primates. It was discovered that they have a high-frequency limit of auditory sensitivity of approximately 91 kHz and are also able to vocalize with a dominant frequency of 70 kHz. These values are an example of ultrasonic communication and are among the highest known for terrestrial mammals. Philippine tarsiers presumably utilize this ability as a private covert communication channel that is undetectable by predators, prey and competitors [6] (F2).

*2.3 Phase III – Launching attacks*

During the last phase of the typical attack scenario an established botnet is utilized to perform different malicious actions, like: launching DDoS attacks, organizing and executing SPAM campaigns, hosting phishing websites, stealing users' credentials, etc..

In this subsection we shortly revise attack techniques, like SPAM campaigns and DDoS, and then provide their equivalents in nature, taking into account their main characteristic features.

**SPAM campaigns.** The emitting of a multitude of unsolicited emails has as its main purpose advertisement. The common denominator of all SPAM is that it is unwanted, and usually originates from multiple sources, thus it is difficult to eradicate.

Invasive species undertake a similar strategy. Their goal of proliferating of the environment is achieved by all sorts of mechanisms for efficient reproduction. One species displaying such aggressive behaviour is an inconspicuous flower, Small Balsam (*Impatiens parviflora*), which possesses an ingenious device for gaining dominance. Its seed pouches are pressurised, and once they are ripe they catapult seeds with high speed, at the slightest touch. This effective means of spreading its genes has made *Impatiens parviflora* an unwelcome guest. Its presence has negative effect on biodiversity, which placed it on some countries' blacklists of species whose cultivation is prohibited.

Likewise, legal steps are also undertaken to restrict the flood of unwanted messages circulating in the Internet. Hosts known to issue large quantities of SPAM are blacklisted, just as the plant, trying to gain dominance in the environment.

**DDoS attacks.** Distributed Denial of Service attacks' aim is to prevent legit users from accessing a service or a resource by occupying it constantly, thus deeming it unavailable to all interested parties. Such attacks are usually conducted from many sources, to achieve appropriate scale of the directed illicit activity.

Same behaviour can be observed in the plant kingdom. *Pueraria montana*, commonly known as the Kudzu vine or Japanese arrowroot. Due to human intervention it has infested North America, southern Africa and some parts of Europe and central Asia. Taken out of its natural habitat, without natural pests and diseases, Kudzu proved a very noxious plant and requires persistent countermeasures to eradicate completely. It is so effective in its mischief, it was used during the Second World War to quickly conceal objects of military significance.

Japanese arrowroot proliferates its ecosystem with astounding speed (ca. 30 cm per day). Within weeks this aggressive perennial plant, similarly like DDoS attack, can literally choke all other growth, including trees and shrubs, by stealing the precious *resources* – light and nutrients.

**3. Defensive techniques**

Not only offensive techniques have their counterparts in the kingdoms of nature - the same happens in the case of security mechanisms and systems. The most popular defensive measures like firewalls, IDS/IPS systems and honeypots bear the same similarities as in the case of the previously discussed attack techniques.

## 3.1 Firewalls

Presently, almost every computer or network possesses a firewall defending it from unwanted network traffic incoming into the machine or network, while permitting useful inbound and outbound traffic.

In this subsection we show analogies for the following firewalls' features:

**(F1)** ability to protect from the threats of the outside world and

**(F2)** ability to filter the unwanted entities.

Animals and plants, just like computers, possess the defensive capacity against unwanted intruders. Hedgehog's spines, porcupine's quills, turtle's shells or acacia thorns are all meant to discourage unwanted attention (F1).

Firewalls' other feature is the capability to differentiate between what is welcome and what is not. Similar selective mechanism is employed by certain plants, which emit chemical substances into the environment which are capable of moderating the growth capacity of other plants. The so called selective allelopathy phenomenon has been observed, among others, in a Mexican shrub. *Leucaena leucocephala* is known to secrete a toxic amino acid stunting the growth of all surrounding plants but not its own seedlings [9] (F2). The same species increases the yield of rice crops but has negative impact on wheat plants. The selective allelopathy bears much resemblance to the digital firewall, permitting only beneficial neighbours to enter a plant's ecosystem.

## 3.2 IDS/IPS

Compared to firewalls, yet more complex defensive capacity is displayed by Intrusion Detection or Prevention Systems. These systems are fine tuned to match occurring events with patterns corresponding to known types of offensive activity (F1) and, if possible, to react appropriately (F2).

Analogical behaviour is frequently displayed in animal communities, where one representative is positioned on a lookout for approaching predators and warns the others of approaching danger.

A great natural IDS/IPS system has been created by the masked birch caterpillar, or *Drepana arcuata*, which resides under silken cover attached to leaves. Once disturbed by leaf vibrations caused by an approaching predator, a *Podisus*, the caterpillar takes preventive actions. It has been found that *Drepana arcuata* reacts only to vibrations characteristic to its natural enemies, while discriminating all non-relevant signals (e.g. caused by rain or wind). When in danger, the caterpillar displays one of the three deterrent behaviours – it scrubs its leaf with its abdomen, drums it or scratches with its mandibles [7]. All these together have been found to successfully deter attackers.

The close inspection of the masked birch caterpillar's behaviour shows that it is as complex, as that of a digital IDS/IPS system. The initial stage of detecting vibrations is followed by an algorithm matching it with the pattern for predator-related disturbances (F1). Finally *Drepana arcuata* takes preventive measures to fend off the attacker (F2).

### 3.3 HoneyPots

Honeypot systems play a significant role in increasing Internet security by means of "trapping" the intruder in an isolated segment of the network, or an application, and observing his behaviour. Frequently a bait is set up – an apparently ill protected resource or dummy resource is displayed for prey, and later monitored to discover attacker's modus operandi.

Lady's-slipper orchid (*Cypripedium calceolus*) behaves very much like a honeypot, it uses its scent and colour to attract pollinating insects. These are tempted to enter the flower's slipper-shaped pouch, where they remain trapped unless they follow the only exit path, which leads the insect via pollen-bearing stamina. The pollinator is released once it is covered with pollen, or has successfully deposited gametes originating from a different orchid. Such pattern of behaviour is encountered quite frequently in the environment, where symbiotic relations are common. *Cypripedium calceolus* restrains the insect until its objective is fulfilled, just as a honeypot lures the intruder to remain inside the specially crafted artificial environment, until a comprehensive attacker behaviour model can be created.

### 4. Conclusions

When looking at the complex web of interactions, whether talking about Internet users, or wild creatures, one can notice an emerging pattern. Where conflict of interest is present, the reason behind it is either gaining competitive dominance over counterparts, or obtaining access to a limited resource. This leads to a situation, where known interaction models are constantly adapted to devise new means of outwitting the rival. Most of the discussed scenarios rely on the inherent reaction model of an entity and its inability to react appropriately if a victim or an attacker exploits this bias for certain signals. Therefore, it becomes apparent that two significant components of providing security are the ability to respond to dubious signals with cautiousness and to possess the knowledge of the behavioural pattern of the aggressor.

The ongoing evolution of the offensive and defensive techniques from animal, plant, or other eukaryote kingdoms finds analogy in network attacks and their countermeasures. Indeed, in both cases we are witnessing an arms race, however taking place in different time windows. For every offensive technique that has been developed, sooner or later, a defensive scheme appears both in the world of IT security as well as in nature.

In the natural environment many predator-prey systems have been experiencing an arms race. For example, many molluscs, such as Murex snails, have developed thick shells to avoid being eaten by animals such as crabs and fish. In turn, predators like crabs have grown more powerful claws and jaws that compensate for the snails' thick shells.

In the Internet malware is developing at such a rate that an "arms race" has ensued between cyber criminals and those seeking to thwart their activities. Malware authors are getting better at being stealthy and finding ways to fight back against the security pros. Therefore, even if a countermeasure has been developed for one type of malware, the attackers usually slightly alter the code to give it a "rebirthing suite" – to improve its defences against antivirus programs. And to use it over and over again. An example of this process is Zeus/ZBOT malware which was originally discovered in 2007 and defensive systems were upgraded to deal with it, but from this time it was resurrected a few times in different parts of the globe in the period of 2009-2012. This again resembles processes in nature as such arms race will surely not come to an end in the foreseeable future.

Lastly, the not so optimistic conclusion is that, judging from the perpetual contention of offensive and defensive techniques in the kingdoms of living things, which hasn't brought around any definite countermeasure, one can expect that IT security will follow the same pattern. As in nature, defensive systems in the virtual world are upgraded only when a new threat has been identified. Typically, such systems do not attempt to foresee potential new strategies or means of attacks to prevent them in advance. Thus, maybe it is now time for researchers and security experts looking out for new network attack techniques and novel defence systems to take a peek at the goings-on in the kingdoms of living things in search of new inspirations…

**Short authors bios**

**Wojciech Mazurczyk** holds an M.Sc. (2004), a Ph.D. (2009, with honours) and and D.Sc. (habilitation, 2014) all in telecommunication from Faculty of Electronics and Information Technology of Warsaw University of Technology, Poland; associate professor at WUT; author of over 80 scientific papers, 1 patent application and 30 invited talks on information security and telecommunications; main research interests: information hiding techniques, network anomalies detection, digital forensics and bio-inspired security and networking. Since 2013 an IEEE Senior Member.

**Elżbieta Rzeszutko** is a research assistant at the Institute of Telecommunications, Faculty of Electronics and Information Technology of Warsaw University of Technology, Poland. She holds and M.Sc. and B.Sc. in telecommunications (both with honours) and is currently a Ph.D. student.

**Contact information**

*Wojciech Mazurczyk*

Email: wmazurczyk@tele.pw.edu.pl

Phone: +48 22 234 6189

Mailing address:

Warsaw University of Technology

Faculty of Electronics and Information Technology

Institute of Telecommunications

15/19 Nowowiejska Str.

00-665 Warszawa, Poland

*Elżbieta Rzeszutko*

Email: E.Rzeszutko@tele.pw.edu.pl

Phone +48 22 234 5894

Mailing address:

Warsaw University of Technology

Faculty of Electronics and Information Technology

Institute of Telecommunications

15/19 Nowowiejska Str.

00-665 Warszawa, Poland